\documentclass{ifacconf}

\usepackage{subcaption}
\usepackage{graphicx}      
\usepackage{natbib}        
\usepackage{amsmath}
\usepackage{amssymb}
\usepackage{bm}
\usepackage[dvipsnames]{xcolor}
\usepackage{csquotes}
\usepackage{url} 

\renewcommand{\url}[1]{} 
\usepackage{siunitx}
\newcommand{\pd}{\partial}
\newcommand{\tdSubScript}{\mathrm{t}} 
\newcommand{\trSubScript}{\mathrm{r}} 

\newcommand*{\IR}[1][]{\mathbb{R}^#1}

\newcommand*{\sv}[1]{\mathsf{#1}}

\newcommand*{\st}[1]{\bm{#1}}

\newcommand*{\LG}{\mathsf{G}}
\newcommand*{\Lg}{\mathfrak{g}}

\newcommand*{\SO}[1]{\mathsf{SO(#1)}}

\newcommand*{\SE}[1]{\mathsf{SE(#1)}}
\newcommand*{\se}[1]{\mathfrak{se}(#1)}
\newcommand*{\dse}[1]{\mathfrak{se}(#1)^*} 

\DeclareMathOperator{\Ad}{Ad}

\newcommand{\rttirm}{\mathrm{d}\tau^{-1}}

\begin{document}
\begin{frontmatter}

\title{Discrete Geometric Modeling and Extended State Estimation of Continuum Robots}

\thanks[footnoteinfo]{Authors contributed equally.}

\author[TUM-RT]{Maximilian Herrmann\thanksref{footnoteinfo}}
\author[TUM-RT]{,  Leander Pfeiffer\thanksref{footnoteinfo}}
\author[TUM-RT]{,  Paul Kotyczka}

\address[TUM-RT]{Technical University of Munich, TUM School of Engineering and Design, Munich Institute of Robotics and Machine Intelligence (MIRMI), Chair of Automatic Control.\\ (e-mail: \{maximilian.herrmann,leander.pfeiffer,kotyczka\}@tum.de).}

\begin{abstract}                
In this paper,
we present a fully discrete approach for the accurate and numerically efficient dynamical modeling and state estimation of continuum robots.
The model is based on geometrically exact beams in a minimal, strain-based formulation and derived in the framework of Lie group variational integrators, allowing to preserve important geometric properties that we exploit to achieve high accuracy and numerical efficiency.
We then propose a disturbance observer based on an extended Kalman filter formulation that reliably estimates system states as well as model uncertainties and external disturbances.
Experiments on a real system validate the accuracy and efficiency of the proposed model and observer.
\end{abstract}


\end{frontmatter}

%
%
%
%
%

\section{Introduction}

Continuum manipulators have attracted widespread interest in the robotics community in the past two decades,
as they promise to enable a wide range of new applications.
Examples are a variety of medical applications, such as minimally invasive surgery or endoscopic procedures \citep{BurgnerKahrs.2015},
safety-critical human-robot interaction scenarios, or inspection tasks in inaccessible, cluttered environments, such as turbines \citep{BDM+21}.

\begin{figure}
   \centering
   \hspace{0.4cm}
   \includegraphics[height=4.7cm]{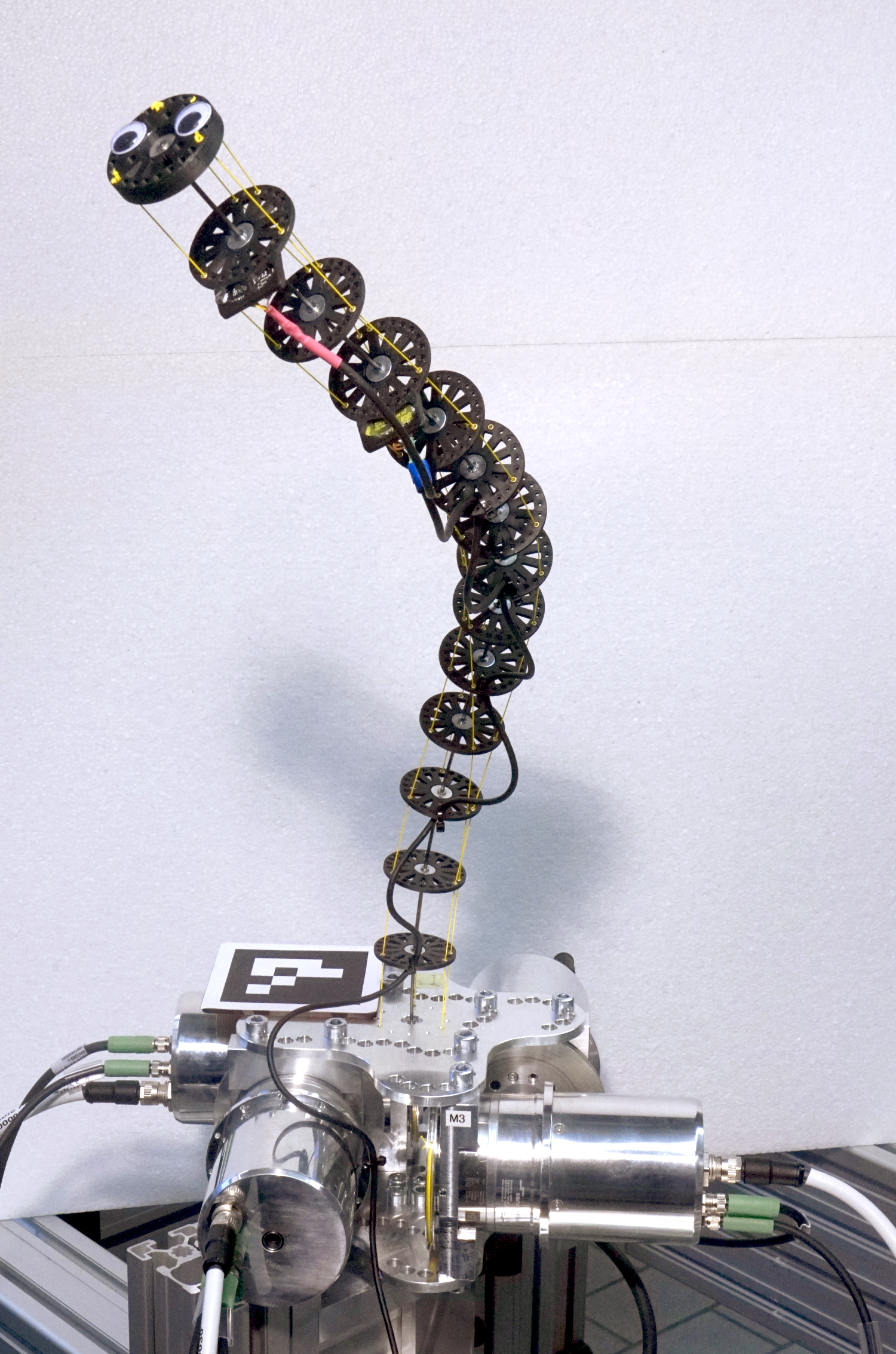}
   \def\svgwidth{59mm}
   \hspace{0.2cm}
   \resizebox{!}{4.7cm}
   {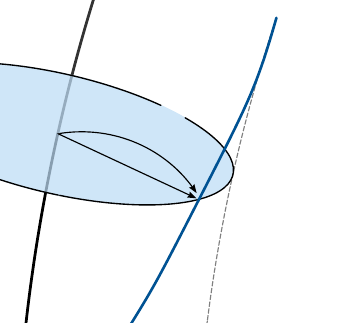\hspace*{-0.4cm}}\hfill
   \vspace*{-0.15cm}
   \caption{\small
      Left: The tendon-driven continuum robot \textsc{Peter} (\emph{\textbf{p}ro-grammable \textbf{e}lastic \textbf{t}endon-\textbf{e}xcited \textbf{r}obot}) with four torque-controlled drives, a spring-steel backbone, spacer discs, and IMUs.
      Right: Tendon actuation kinematics.}
   \label{fig_photo_peter}
\end{figure}

Key element of this class of \emph{soft robots} is usually a highly elastic \emph{backbone} that can deform continuously over its length, Fig.~\ref{fig_photo_peter}.
While many different actuation methods have been proposed and investigated, one of the most versatile types is tendon actuation, in which several tendons are routed along the backbone;
applying specified tensions to the tendons exerts a distributed force on the backbone,
causing a specific, possibly highly complex deformation.

A central challenge in realizing complex tasks remains the dynamical modeling of these systems.
While essential for efficient numerical simulation and model-based control,
accurate models are often high-dimensional, nonlinear and numerically stiff, making time integration difficult with standard methods.
Often, strong limitations in terms of accuracy and generality must be accepted to achieve the low computation times required by applications in robotic real-time contexts.

\graphicspath{{figures}}

A variety of modeling approaches have been proposed to address these challenges.
An increasingly popular, physically consistent approach is to model the backbone using (geometrically exact) Cosserat beams,
which were first used by \cite{Rucker2011ContRobotsGeneralTendonRouting} for fully dynamical models of tendon-driven continuum robots (TDCRs).
These can accurately describe arbitrarily large static and dynamic deformations caused by external forces, including gravity and contact forces.
While very general, they also result in increased complexity and possibly high-dimensional models,
and require careful spatial discretization to obtain numerically efficient and practically usable models, see e.g., \cite{Renda2018DiscreteCosseratApproach, Boyer.2020}.
Strain-based discretizations using specialized shape functions such as the Cosserat model of \cite{Renda.2020, Renda.2022} achieve very low-dimensional and efficient models; however, they only remain accurate in the absence of external forces (including gravity), making them unusable for many practical scenarios.

Many of these works incorporate the inherent Lie group structure of Cosserat beams,
whose configuration space is an infinite-dimensional product of the special Euclidean group $\SE3$.
So far, however, this property has never been explicitly preserved in space and time discretization to improve the numerical efficiency of discrete TDCR models.
Exactly this can be elegantly achieved in the framework of discrete mechanics and (Lie-group) variational integrators \citep{Marsden.2001, Lee2008CompGeomMechDiss},
where a discrete-time model is directly derived based on discrete analogues of continuous-time variational principles,
preserving important geometric structures like symmetries, symplecticity, or the (Lie-group) configuration space structure.
The discrete models have favorable numerical advantages: They are symplectic, momentum preserving,
and stable at large time steps.
In this variational Lie-group framework, \cite{HK24} proposed a fully discrete, strain-based model of geometrically exact beams that is accurate, numerically efficient, and structure-preserving. 
The specific strain-based (relative kinematic) formulation uses relative deformations as configuration variables,
allowing the constraint-free exclusion of stiff deformation modes such as shear and elongation,
which is crucial for efficient time integration. In this paper, we extend this approach to a fully discrete model for TDCRs by additionally incorporating external masses and tendon actuation in the same structure-preserving framework.
The discrete Lie-group model can provide accurate, physically consistent solutions even at coarse space and time discretizations,
making it numerically highly efficient.

To practically realize complex controller structures involving state feedback,
accurate information about the robot's states is required.
Since these usually correspond to quantities like discrete strains (as in our model), or configurations of discrete beam nodes, direct measurement is often difficult.
This requires using state observers, which must run in real time while providing accurate estimates even if model inaccuracies and external forces are present.

\cite{LBB22,LBB24} and \cite{TLB25} employed Gaussian process regression in combination with Cosserat rod and Lie-group formulations to fuse strain and pose feedback. These observers demonstrated accurate state estimation under nominal conditions but require known or negligible external disturbances. Similarly, \cite{FMA+24} and \cite{ZHL24} proposed Riccati and boundary observer designs on Lie groups, proving local observability and asymptotic convergence under idealized, disturbance-free assumptions.

Several works have begun to treat external disturbances as forces to be estimated alongside the system state. For quasi-continuum manipulators, \cite{MVS20} and \cite{VMS21} introduced extended state observers ({ESOs}) that augment the system state with lumped disturbance terms, showing improved tracking and robustness. In the context of TDCRs, \cite{ADBR22} proposed disturbance observers based on extended Kalman filters (EKFs) using a simplified geometric model. A recent approach by \cite{FRW24} combined state and disturbance estimation using Gaussian processes, though only for static deformations and without real-time capability.

The rest of the paper is organized as follows:
In Section 2 of this paper, we introduce the variational, fully discrete TDCR model based on the aforementioned discrete Cosserat model. As a new contribution,
we extend it with tendon actuation with arbitrary paths in the same structure-preserving framework.
Based on this dynamic model, an observer design is then proposed in Section 3, which builds upon a symplectic EKF formulation to accurately estimate both states and disturbances of the manipulator.
Finally, in Section 4, we validate the model and observer on the real-world experimental robot depicted in Fig.~\ref{fig_photo_peter},
showing excellent agreement with experimental data.



\section{System Modeling}
\label{sec:system_modeling}
In this section, we extend the pure beam model of \cite{HK24} to a discrete TDCR model.
We first introduce the kinematics and the resulting equations of motion; for the derivation and all further details, however, we refer to the original paper for brevity.
As a new result, we then derive the extension with tendon actuation in the same structure-preserving framework.
Note that the model heavily relies on Lie group concepts and the associated notation conventions,
which unfortunately cannot be properly introduced here,
and we again refer to the full paper for details and references.


The instantaneous configuration of a geometrically exact Cosserat beam (see, e.g., \cite{Antman.2005}) with length $L$ is defined by its line of centroids, given by the map
$x : [0, L] \to \IR3$ with respect to an inertial frame,
and the orientations of its infinitesimal cross-sections along this center line.
The orientation of a body-fixed orthonormal cross-section frame at the arc length $s\in [0,L]$ with respect to the inertial frame is determined
by the rotation matrix $R : [0, L] \to \SO3$,
defining the cross-section pose by the map $g : [0, L] \to \SE3$, where $g(s) = (R(s), x(s))$.
The derivatives of the configuration in space and time are given by the left-trivialized (body-fixed) cross-section velocity
\begin{equation}\label{eq_def_eta_beam}
   \hat{\eta}(s)
   =
   g(s)^{-1} \dot{g} (s)
   \in \se3
\end{equation}
and the left-trivialized (body-fixed) deformation gradient
\begin{equation}\label{eq_def_xi}
   \hat{\xi}(s)
   =
   g(s)^{-1} g'(s)
   \in \se3,
\end{equation}
where $\dot{(\,\cdot\,)}$ denotes the time derivative and $(\,\cdot\,)'$ the spatial derivative along $s$.
Defining the usual kinetic, strain and potential energy densities based on these quantities leads to the fully continuous (partial-differential) beam model.

For space and time discretization, we divide the beam in $n$ discrete segments and introduce the sequence of time $\{t^k = k h \mid k = 0, \dots, N \}$.
We use a Lie-group structure-preserving discretization with discrete Lie-algebra approximations $\hat{\eta}_a^k$ and $\hat{\xi}_a^k$ of the continuous velocity and deformation gradient computed via the \emph{retraction map} (here, the Cayley map).
The segment-wise \emph{discrete deformations} $\hat{\xi}_a^k$ represent discrete strains and are used as a relative parameterization of the beam kinematics.
Finally, the fully discrete dynamical model is obtained by applying the \emph{discrete Lagrange-d'Alembert principle} to a \emph{discrete Lagrangian} that is defined by using the (generalized) trapezoidal rule in space and time.
This results in the discrete equations
\begin{multline}\label{eq_DEL_structured}
   \sum_{a=1}^{n}
   ({J}_a^k)^T
   \left(
   \tfrac{1}{h}\mu_{a}^k
   - \tfrac{1}{h} \Ad_{\tau(h\hat{\eta}_a^{k-1})}^* \mu_a^{k-1}
   + v(g_{a}^k)
   - f_a^k
   \right)
   \\
   + \sv{K} (\sv{q}^k - \bar{\sv{q}})
   + \sv{D}\dot{\sv{q}}^k
   - \sv{B}(\sv{q}^k)\sv{u}^k
   = 0,
\end{multline}
where ${J}_a^k \in \IR{6 \times n\cdot r}$ is the \emph{geometric Jacobian} of node $a$ at time step $k$,
$v(g_{a}^k)$ encapsulates the effects due to gravity, $\hat{f}_a^k$ is an external, body-fixed wrench, and
\begin{equation*}\label{eq_mu_a_k}
   \mu_{a}^k (\eta_{a}^k)
   =
   \left( \rttirm_{h\hat{\eta}_{a}^k} \right)^*
   \mathbb{M}_a \eta_a^k,
   \qquad
   \hat{\mu}_{a}^k \in \dse{3}
\end{equation*}
is the \emph{discrete momentum} with the generalized node inertia matrix $\mathbb{M}_a \in \IR{6\times6}$ and the right-trivialized derivative of the inverse retraction map $\rttirm$.
$\sv{K}$ and $\sv{D}$ represent the system stiffness and dissipation matrices, respectively,
and $\sv{q} \in \IR{n\cdot r}$ is the configuration vector consisting of the allowed discrete deformations of all segments.
The $r$ allowed deformation modes can be specifically selected; for example, the commonly used Kirchhoff beam with $r=3$ only allows bending and torsion modes,
eliminating the highly stiff, numerically problematic shear and elongation modes.
The derivative $\dot{\sv{q}}^k$ is approximated via finite differences; for the implementations in this paper, we use a first-order approximation.
%
%
Finally, 
$\sv{B}(\sv{q}) \in \IR{n\cdot r \times p}$ is a general input matrix and $\sv{u}^k = \sv{u}(t^k) \in \IR{p}$ are the corresponding inputs that we use here to incorporate tendon actuation.

As with most variational integrators, the nonlinear DEL equations \eqref{eq_DEL_structured} are implicit, and must be solved numerically for the unknown configurations $\sv{q}^{k+1}$ (contained in $\mu^k_a$ and $\dot{\sv{q}}^k$) in each time step.
Since the implicit part is not complicated, this can be done very efficiently; here, we use Broyden's good method.
Overall, we obtain a highly efficient model that is very well-suited for time integration of stiff, complex systems such as the present TDCR.


Next, we extend the derived model by tendon actuation.
While the resulting (spatially continuous) forcing term is equivalent to the derivations in \cite{Rucker2011ContRobotsGeneralTendonRouting}, where tendon actuation with arbitrarily routed, not necessarily straight tendons was considered for the first time, our derivation is based on the virtual work principle as in \cite{Boyer.2020},
which directly fits into the variational derivation of the model above.

We consider $p$ tendons routed in arbitrary paths along the backbone,
although we initially only focus on one tendon to simplify notation.
We assume that the tendons can only transmit force in their tangential direction and have negligible mass,
and, for now, that they can move frictionless (in practice, however, tendon friction can be significant and complex; this is, however, outside the scope of the present paper).
The path of the considered tendon is determined (Fig. 1, right) by the position vector $x_{\trSubScript} : [0,L] \rightarrow \IR3$ defining the tendon position relative to the backbone at the arc length $s$ in the local body-fixed cross-section frame (hence, its $z$ component is always zero).
Accordingly, we can define the associated transformation
$ 
   g_{\trSubScript}
   =
   (R_{\trSubScript}, x_{\trSubScript}) : [0, L] \rightarrow \SE3,
$ 
where $R_{\trSubScript} \in \SO3$ is a rotation matrix that defines the orientation of a local frame located in the tendon path, which has its $z$-axis aligned with the tendon path's tangent vector similar to the  Frenet-Serret convention.
Then, the absolute tendon path w.r.t. the inertial frame is
\begin{equation}\label{eq_cablePathSE3}
   g_{\tdSubScript} (s)
   =
   (R_{\tdSubScript}(s), x_{\tdSubScript}(s))
   =
   g (s) \, g_{\trSubScript} (s) \in \SE3
\end{equation}
with the associated left-trivialized deformation gradient
\begin{equation*}
   \hat{\xi}_{\tdSubScript}
   =
   g_{\tdSubScript}^{-1} g_{\tdSubScript}' \in \se3,
   \qquad
   \xi_{\tdSubScript}
   =
   \begin{bmatrix}
      \Omega_{\tdSubScript}^T &
      \Gamma_{\tdSubScript}^T
   \end{bmatrix}^T
   \in \IR6,
\end{equation*}
where the translational component $\Gamma_{\tdSubScript} =
R^T x_{\tdSubScript}' \in \IR3$ corresponds to the tendon path's tangent vector in the local body-fixed frame.
Furthermore, the length of a tendon terminated at the arc length $L_{\tdSubScript} \in [0, L]$ is
\begin{equation}\label{eq_tendonLengthCont}
   l_{\tdSubScript}
   =
   \int_{0}^{L_{\tdSubScript}}
   \left \| x_{\tdSubScript}' \right \|
   \mathrm{d}s
   =
   \int_{0}^{L_{\tdSubScript}}
   \left \| \Gamma_{\tdSubScript} \right \|
   \mathrm{d}s.
\end{equation}
Noting that the variation of the tangent vector $\delta \Gamma_t$ can be expressed as
\begin{equation*}
   \delta \Gamma_{\tdSubScript}
   = \delta ( \Gamma_{\trSubScript} + \hat{\Omega} x_{\trSubScript} + \Gamma)
   = \delta \hat{\Omega} x_{\trSubScript} + \delta\Gamma,
\end{equation*}
we can compute the (infinitesimal) variation of the tendon length as (cf. \cite{Boyer.2020})
\begin{align*}
   \delta l_{\tdSubScript}
   =
   \int_{0}^{L_{\tdSubScript}}
   \frac{1}{\left\| \Gamma_{\tdSubScript} \right \|}
   \begin{bmatrix}
      \hat{x}_{\trSubScript} \Gamma_{\tdSubScript} \\ \Gamma_{\tdSubScript}
   \end{bmatrix}
   \cdot
   \delta \xi
   \,
   \mathrm{d}s,
\end{align*}
and the virtual work done by the actuated tendon is
\begin{equation}\label{eq_virtualWorkCableCont}
   \delta W_{\tdSubScript}
   =
   u(t) \delta l_{\tdSubScript}
   =
   \int_{0}^{L_{\tdSubScript}}
   \frac{u(t)}{\left\| \Gamma_{\tdSubScript} \right \|}
   \begin{bmatrix}
      \hat{x}_{\trSubScript} \Gamma_{\tdSubScript} \\ \Gamma_{\tdSubScript}
   \end{bmatrix}
   \cdot
   \delta \xi
   \,
   \mathrm{d}s,
\end{equation}
where $u: \IR{} \rightarrow \IR{+}$ is the (positive) tendon tension.
Applying the Lagrange-d'Alembert principle, we directly obtain the distributed force term (force density)
\begin{equation}\label{eq_cableForceCont}
   f_{\tdSubScript}(t,s)
   =
   \frac{u(t)}{\left\| \Gamma_{\tdSubScript} \right \|}
   \begin{bmatrix}
      \hat{x}_{\trSubScript} \Gamma_{\tdSubScript} \\ \Gamma_{\tdSubScript}
   \end{bmatrix}
   \in \IR6
\end{equation}
for one tendon,
which can be understood as a distributed inner force analogous to the stress and dissipation terms.

For spatial discretization in the same structure-preserving framework as above,
we first define the discrete deformation gradient of the tendon path along a beam segment $a$ with length $l$ in analogy to the discretization shown in \cite{HK24} as
\begin{equation*}
   \hat{\xi}_{\tdSubScript,a}
   =
   \tau^{-1} \left(  (g_{\tdSubScript,a})^{-1} g_{\tdSubScript,a+1} \right) / \, l \in \se3,
   \;
   \xi_{\tdSubScript,a}
   =
   \begin{bmatrix}
      \Omega_{\tdSubScript,a}\\
      \Gamma_{\tdSubScript,a}
   \end{bmatrix} \in \IR6,
\end{equation*}
where we assume that the length of the tendon segment remains approximately constant in the deformed case.
By substituting \eqref{eq_cablePathSE3}, it can be rewritten as
\begin{equation}\label{eq_xi_t_a}
   \hat{\xi}_{\tdSubScript,a}
   =
   \tau^{-1} \left(  g_{{\trSubScript},a}^{-1} \, \tau(l \xi_a) \, g_{{\trSubScript},a+1} \right) / \, l.
\end{equation}
Consistent with the beam discretization, we approximate the integral in \eqref{eq_virtualWorkCableCont} using the trapezoidal rule and obtain
\begin{multline*}
   \delta W_{\tdSubScript}
   =
   \int_{0}^{L_{\tdSubScript}}
   f_{\tdSubScript}(s)
   \cdot
   \delta \xi
   \,
   \mathrm{d}s
   \approx
   \sum_{a=0}^{n-1}
   \tfrac{l}{2}
   \left(
   f_{\tdSubScript}(s_{a})
   +
   f_{\tdSubScript}(s_{a+1})
   \right)
   \cdot
   \delta \xi_{a}
   \\
   =
   \sum_{a=0}^{n-1}
   \frac{u(t) l}{\left\| \Gamma_{\tdSubScript,a} \right \|}
   \begin{bmatrix}
      \frac{1}{2}\left( \hat{x}_{{\trSubScript},a} + \hat{x}_{{\trSubScript},a+1}\right) \Gamma_{\tdSubScript,a}
      \\
      \Gamma_{\tdSubScript,a}
   \end{bmatrix}
   \cdot
   \delta \xi_{a}.
\end{multline*}
From this equation, we directly obtain the discrete tendon force on a {segment} $a$
\begin{equation*}
   f_{\tdSubScript,a} (t)
   =
   b_a(\xi_a) \, u(t)
\end{equation*}
written in terms of the segment actuation vector
\begin{equation}\label{eq_b_a}
   b_a(\xi_a)
   =
   \frac{l}{\left\| \Gamma_{\tdSubScript,a} \right \|}
   B_a^T
   \begin{bmatrix}
      \frac{1}{2}\left( \hat{x}_{{\trSubScript},a} + \hat{x}_{{\trSubScript},a+1}\right) \Gamma_{\tdSubScript,a}
      \\
      \Gamma_{\tdSubScript,a}
   \end{bmatrix} \in \IR{r},
\end{equation}
where we additionally included the selection matrix $B_a$, which limits the allowed deformation modes as introduced in \cite{HK24}. We can then directly assemble the input matrix $\sv{B}(\sv{q}) \in \IR{n\cdot r \times p}$ in \eqref{eq_DEL_structured} composed of the individual actuation vectors \eqref{eq_b_a}, given that $f_{\tdSubScript,a} (t) = \sum_{i=1}^{p} u_i(t) \, b_{a,i}(\xi_a)$.

\newpage

\section{State and Disturbance Estimation}
\label{sec_observer_design}

Our proposed observer design builds on the EKF, but extends it with respect to the underlying model, where we leverage the symplectic integration step defined by the previously introduced model.
Given that the state and disturbances of continuum manipulators are tightly coupled (\cite{FRW24}) we propose an observer design with an extended state formulation. Extending the state vector $\vec{x}^k$ with the external disturbance vector $\vec{\phi}^k_{\mathrm{ext}}$ leads to the discrete state definition
$ \vec{x}^k_{\mathrm{ext}} = \begin{bmatrix} (\sv{q}^k)^T &  (\sv{q}^{k-1})^T &  (\vec{\phi}^k_{\mathrm{ext}})^T\end{bmatrix}^T.$
The disturbance vector
encapsulates external forces and forces originating in modeling uncertainties.
Given the geometry of the manipulator, we assume only external forces and no external moments. While we could select $\vec{\phi}^k_{\mathrm{ext}}$ to represent all forces acting on each discretized frame of the manipulator, we limit $\vec{\phi}^k_{\mathrm{ext}}$ to the forces acting on the final frame ($a = n$), given that external contact is most probable here.
We assume piece-wise constant disturbances ($\vec{\phi}^k_{\mathrm{ext}} = \vec{\phi}^{k+1}_{\mathrm{ext}}$), similar to the approaches taken by \cite{MVS20} and \cite{VMS21}.

The EKF operates in two stages; prediction and correction. The index $(\cdot)^{k|k+1}$ denotes a predicted value at $k+1$ based only on information from time step $k$,
and we denote observed values with $\Bar{(\cdot)}$.

\subsection{Prediction Step}
\label{ssec:prediction_step}
We use the relative-kinematic beam model by solving \eqref{eq_DEL_structured} to predict the next state $\sv{\Bar{q}}^{k|k+1}$ using the current estimate $\vec{\bar{x}}^k_{\mathrm{ext}}$, which serves as an efficient prediction that preserves symplecticity. A similar approach for a Hamiltonian system formulation is applied in \cite{FM15, SFM17} using the symplectic Euler method. However, given the complexity of \eqref{eq_DEL_structured}, their approach to computing the discrete system Jacobian $\bm{A}^k$ is not feasible for the present case, as a direct one-step map would need to be derived.

For the Jacobian $\bm{A}^k$, it is central that the discrete state update equation \eqref{eq_DEL_structured} is linearized around the current state, rather than linearizing the continuous system dynamics \citep{MJ11}. We can thus state that \begin{equation}
	\bm{A}^k  = \left. \frac{\pd \bm{x}^{k+1}_{\mathrm{ext}}}{\pd \bm{x}^k_{\mathrm{ext}}} \right|_{\bm{\Bar{x}}^k_{\mathrm{ext}}} = \left. \begin{bmatrix}	\frac{\pd \sv{q}^{k+1} }{\pd \sv{q}^k} & \frac{\pd \sv{q}^{k+1} }{\pd \sv{q}^{k-1}} & \frac{\pd\sv{q}^{k+1} }{\pd\bm{\phi}^k_{\mathrm{ext}}} \\ \bm{I} & \bm{0} & \bm{0} \\ \bm{0} & \bm{0} & \bm{I} \end{bmatrix} \right|_{\bm{\Bar{x}}^k_{\mathrm{ext}}}.
\end{equation}
Given that $\sv{q}^{k+1}$ results from solving \eqref{eq_DEL_structured}, which will be denoted as $\mathrm{DEL} = 0$ in the following, we leverage the implicit function theorem to compute the specific Jacobians with
\begin{align*}
	\frac{\pd \sv{q}^{k+1} }{\pd \bm{w}} &= -\left(\tfrac{\pd \mathrm{DEL} }{\pd \sv{q}^{k+1}}\right)^{-1} \tfrac{\pd \mathrm{DEL} }{\pd \bm{w}}, \label{eq:impl_jacobian_qk} \;
	\mathrm{with} \; \bm{w} = \{ \sv{q}^k, \sv{q}^{k-1}, \bm{\phi}^k_{\mathrm{ext}}\}.
\end{align*}
We reuse the predicted state $\sv{\Bar{q}}^{k|k+1}$  and previous estimate  $\vec{\bar{x}}^k_{\mathrm{ext}}$ to compute $\bm{A}^k$. A numerical approximation is viable, but less efficient, given that \eqref{eq_DEL_structured} needs to be solved for every variation of the state. Instead, we deem it most efficient to precompute the analytical derivative using symbolic programming, which can be evaluated at runtime.
As the last step of the prediction step, the error covariance can be predicted analogously to the classical {EKF} as $\bm{P}^{k|k+1}  =\bm{A}^k \bm{P}^k (\bm{A}^k)^T + \bm{Q}.$

\subsection{Correction Step}
\label{ssec:correction_step}
To correct the prediction step, we need to predict our measurement $\bm{\Bar{y}}^{k|k+1}$.
The available measurements of the system (cf. Sec.~\ref{subsec_exp_system}) can be summarized as  \begin{align*}
	\bm{y}^{k+1} = \begin{bmatrix}
		\bm{\omega}_{\mathrm{IMU},1}^T &\bm{a}_{\mathrm{IMU},1}^T &\bm{\omega}_{\mathrm{IMU},2}^T &\bm{a}_{\mathrm{IMU},2}^T  &\bm{l}_{\tdSubScript}^T
	\end{bmatrix}^T.\end{align*} Given the included acceleration term, we require the predicted state $\sv{\Bar{q}}^{k|k+1}$, along with the previously estimated states $\sv{\Bar{q}}^k$ and $\sv{\Bar{q}}^{k-1}$ to determine the predicted acceleration.
Here, we used finite difference gradients for both accelerations and angular velocities,
with the tendon lengths given by a discretization of \eqref{eq_tendonLengthCont} using \eqref{eq_xi_t_a}.


The Jacobian of the measurement equation $\bm{C}^k$ is similarly precomputed to improve performance, resulting in it being explicitly defined through $\sv{\Bar{q}}^{k|k+1}$, $\sv{\Bar{q}}^{k}$ and $\sv{\Bar{q}}^{k-1}$.
The Kalman gain is computed at every time step as $\bm{K}^k = \bm{P}^{k|k+1} (\bm{C}^k)^T \left(\bm{C}^k \bm{P}^{k|k+1}  (\bm{C}^k)^T + \bm{R} \right)^{-1}$, which is then used to update both the state estimate $$\bm{\Bar{x}}^{k+1}_{\mathrm{ext}} = \bm{\Bar{x}}^{k|k+1}_{\mathrm{ext}} + \vec{K}^k \left( \vec{\tilde{y}}^{k+1} - \vec{\Bar{y}}^{k|k+1} \right)$$ and covariance estimate using the Joseph form $$\bm{P}^{k+1} = (\bm{I} - \bm{K}^k \vec{C}^k)\vec{P}^{k|k+1} (\bm{I} - \bm{K}^k \vec{C}^k)^T + \bm{K}^k \bm{R} (\bm{K}^k)^T.$$

\section{Experimental Validation}
\label{sec_experimental_system}

In this section, we validate the model and the observer on the experimental TDCR \textsc{Peter} shown in Fig. 1 (left).



\subsection{Experimental System}
\label{subsec_exp_system}
Our TDCR consists of a spring-steel backbone with a diameter of $\SI{2}{\milli \metre}$ and a length of $\SI{0.7}{\metre}$. Fourteen spacer disks are mounted via shaft–hub connections at intervals of $\SI{50}{\milli \metre}$,
and each disk guides the tendons through metal grommets at an offset of $\SI{20}{\milli \metre}$ from the backbone.
For distributed sensing, two IMUs have been mounted at disks 11 and 13, where each IMU measures linear acceleration and angular velocity in a body-fixed frame at $\SI{3.33}{\kilo \hertz}$.
The tendons are actuated by four \emph{Sensodrive SensoJoint 3008} actuators, each equipped with high-resolution encoders, an output torque sensor, and output torque control at $\SI{4}{\kilo \hertz}$.
Three of the four tendons are routed in parallel $\SI{120}{\degree}$ apart, and one is routed in a helical path, which is, however, not used in the scope of this paper.
All higher-level real-time control tasks are executed on a \emph{Speedgoat Performance} real-time computer.




\subsection{Model Validation}

We now validate the model by comparing the measured system outputs of static and dynamic experiments to the system outputs from corresponding static and dynamic simulations\footnote{All simulations are implemented with our MATLAB toolbox \textsc{Elara} available under \texttt{https://github.com/ELARA-Toolbox}.};
in the static case, constant input tensions are used, and in the dynamic case, time-varying input trajectories. 
The backbone is modeled as a Kirchhoff beam with $n=12$ segments.

\graphicspath{{plots}}
\begin{figure*}[htp!]
   \centering
   \begin{subfigure}{0.245\linewidth}
      \includegraphics[width=\linewidth]{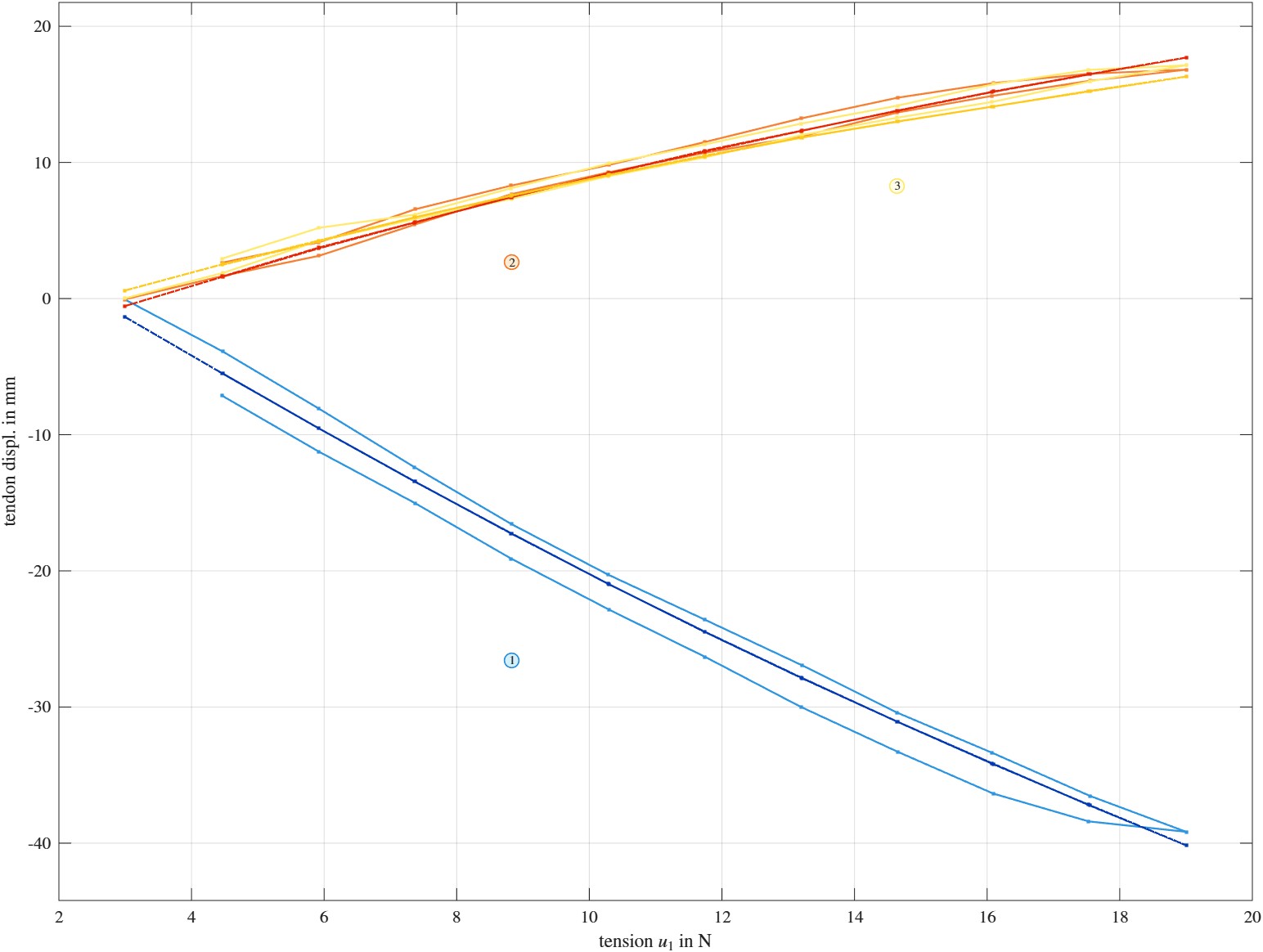}
      \vspace*{-0.4cm}
      \caption{Tendon displacements}
      \label{fig_val_model_static_L}
   \end{subfigure}
   \hfill
   \begin{subfigure}{0.245\linewidth}
      \includegraphics[width=\linewidth]{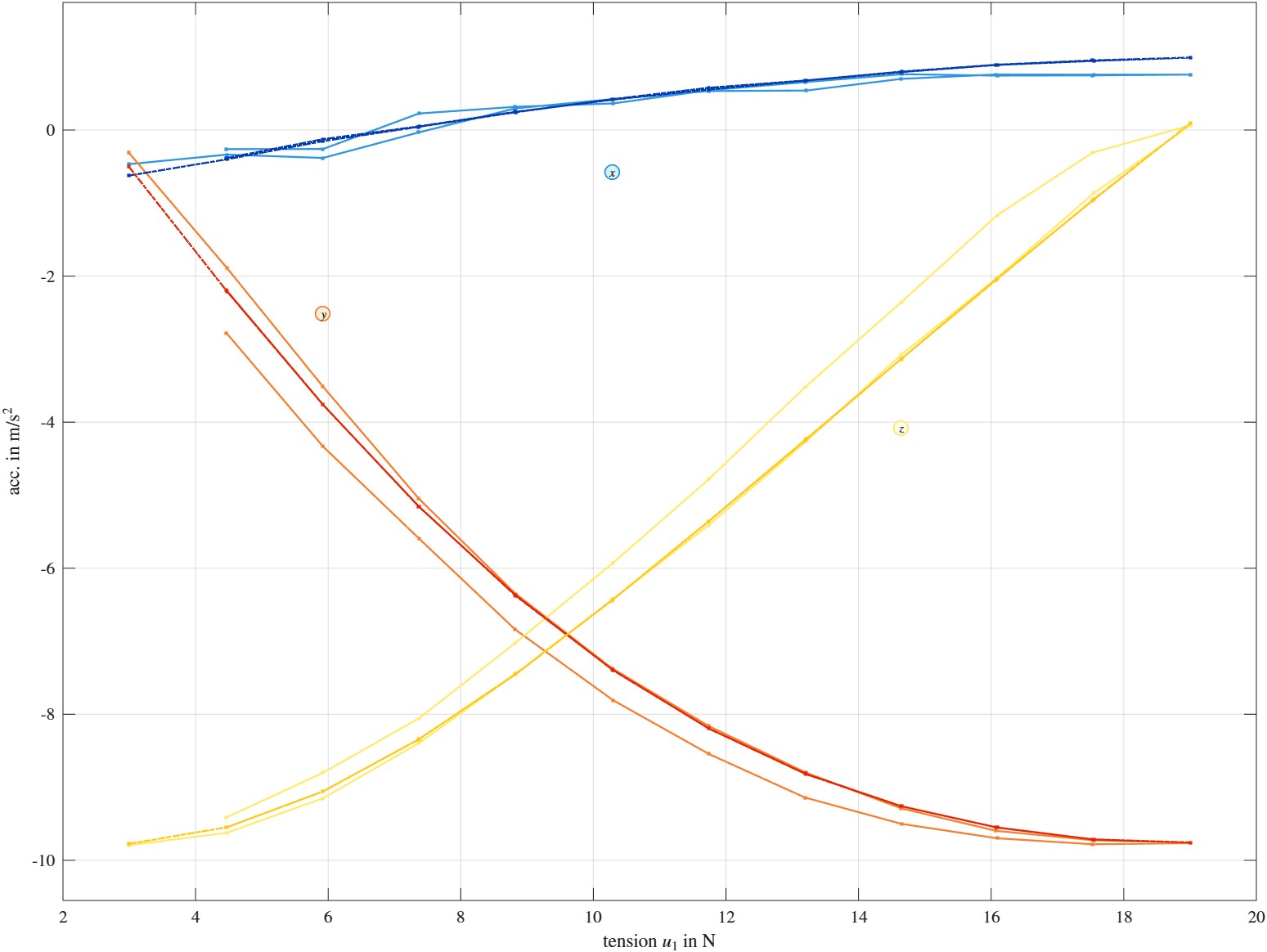}
      \vspace*{-0.4cm}
      \caption{Accelerations of IMU 1}
      \label{fig_val_model_static_IMU1}
   \end{subfigure}
   \hfill
  \begin{subfigure}{0.245\linewidth}
		\includegraphics[trim={0 2cm 0 4cm},clip,width=\linewidth]{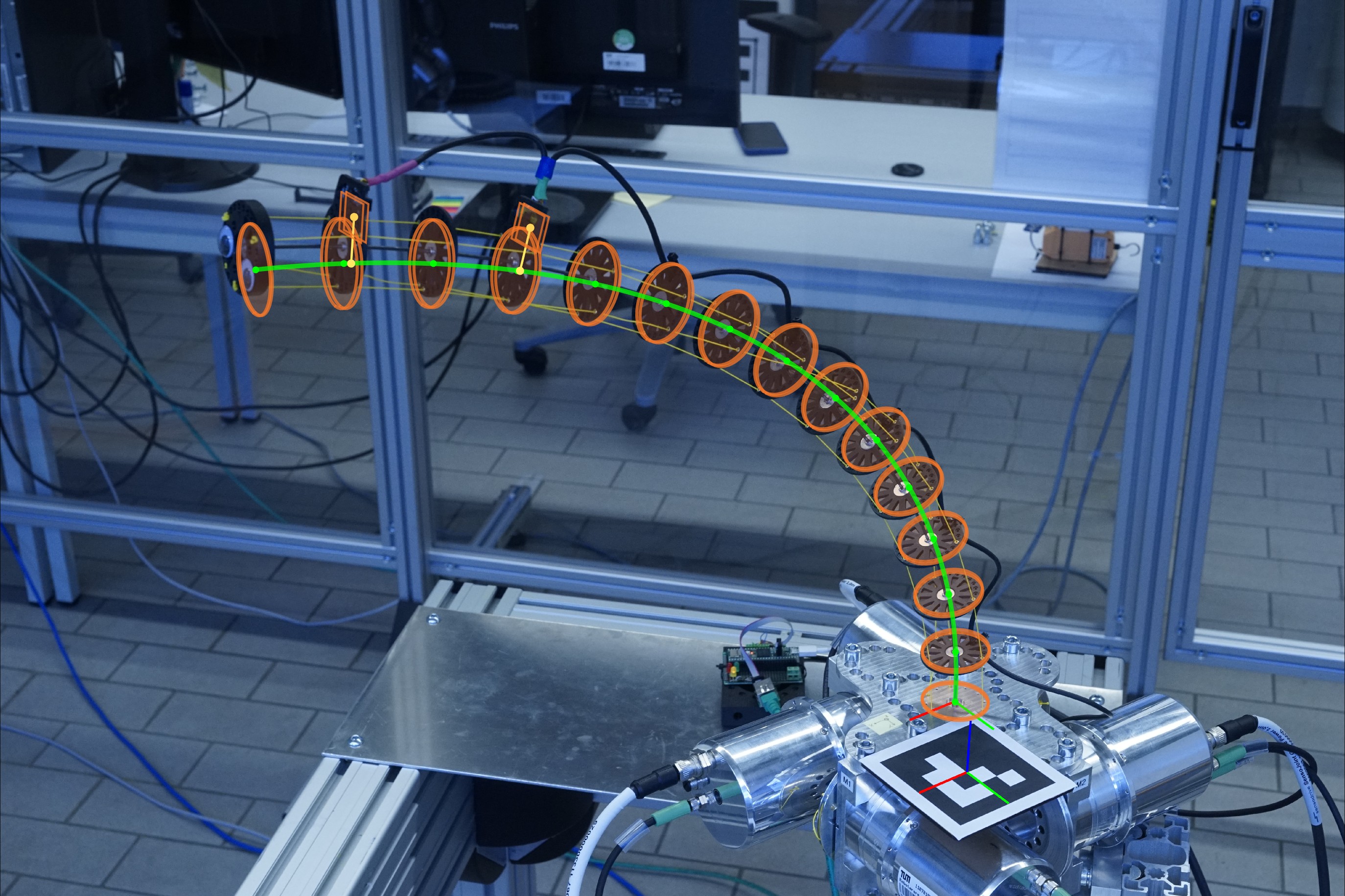}
      \vspace{-0.27cm}
		\caption{Configuration for $u_1 = \SI{11.64}{\newton}$}
		\label{fig_val_model_visual_1}
	\end{subfigure}
	\hfill
	\begin{subfigure}{0.245\linewidth}
		\includegraphics[trim={0 2cm 0 4cm},clip,width=\linewidth]{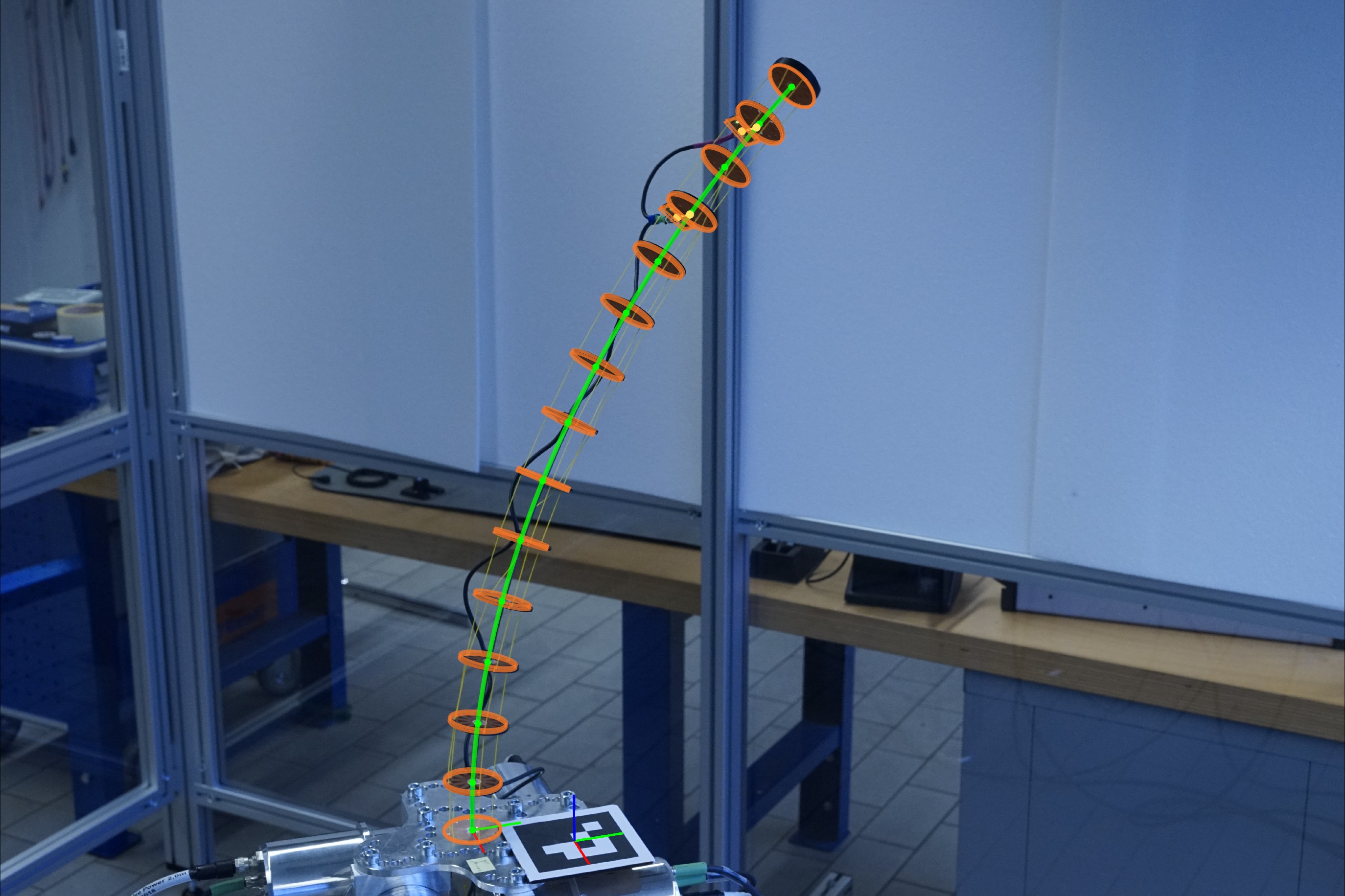}
      \vspace{-0.27cm}
		\caption{Configuration for $u_2 = \SI{7.27}{\newton}$}
		\label{fig_val_model_visual_2}
	\end{subfigure}
   \caption{Results of the static validation: (a) and (b) show the tendon length and acceleration measurements of simulations (dark colors) and the experiment (bright colors),
      where different setpoints of constant tensions are applied to the tendons.
      (c) and (d) show examples of visual validations, where the simulated backbone shape is overlaid on a photo of the robot;
      in addition to the indicated tension, a pretension of \SI{3}{\newton} is applied to all tendons.
   }
   \label{fig_val_model_static}
\end{figure*}

While many of the model parameters are directly available from the robot design,
some parameters can only be estimated, such as the exact backbone stiffness (which is different from the rod alone due to the mounted spacer disks),
the exact inertial properties of the IMUs and the attached cables, and the friction and dissipation properties of the overall system.
Moreover, the mounted IMUs may exhibit slight orientation errors, and the system possesses some additional unmodeled effects such as tendon elasticity and the remaining effective inertia of the torque-controlled drives.
Ideally, static and dynamic parameter identification would be used to obtain a highly accurate model;
however, only slightly tuned nominal parameters are used here (with the IMU orientations being calibrated from experimental data).

In all experiments, a pretension of \SI{3}{\newton} was applied to the straight tendons.
Moreover, as a pragmatic remedy to reduce the influence of static tendon friction (which is significant in the present system), we superpose the commanded tensions with an additional sinusoidal \emph{jitter} signal (\SI{33}{\hertz}, \SI{4.5}{\newton} amplitude) that does not cause a noticeable motion of the overall robot, but significantly reduces stiction and hysteresis.
In the plots shown below, the majority of this signal has been filtered out.
Furthermore, tendon lengths are scaled with identified factors that model linear tendon elasticity.

As an example, Fig.~\ref{fig_val_model_static} (a) and (b) show the results of the static validation for different constant setpoints on the first tendon; the static model matches the experimental results very well.
The results for the other tendons and setpoints are similar.

\begin{figure*}
   \centering
  \begin{subfigure}[t]{0.245\linewidth}
       \includegraphics[width=\linewidth]{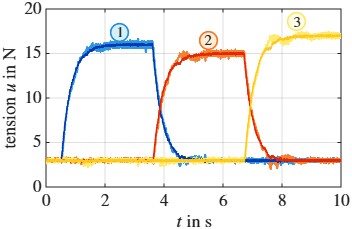}
       \vspace*{-0.5cm}
       {\small\caption{
       Tendon tensions
       }}
       \label{fig_val_model_dyn_tensions}
   \end{subfigure}
   \hfill
   \begin{subfigure}[t]{0.245\linewidth}
      \includegraphics[width=\linewidth]{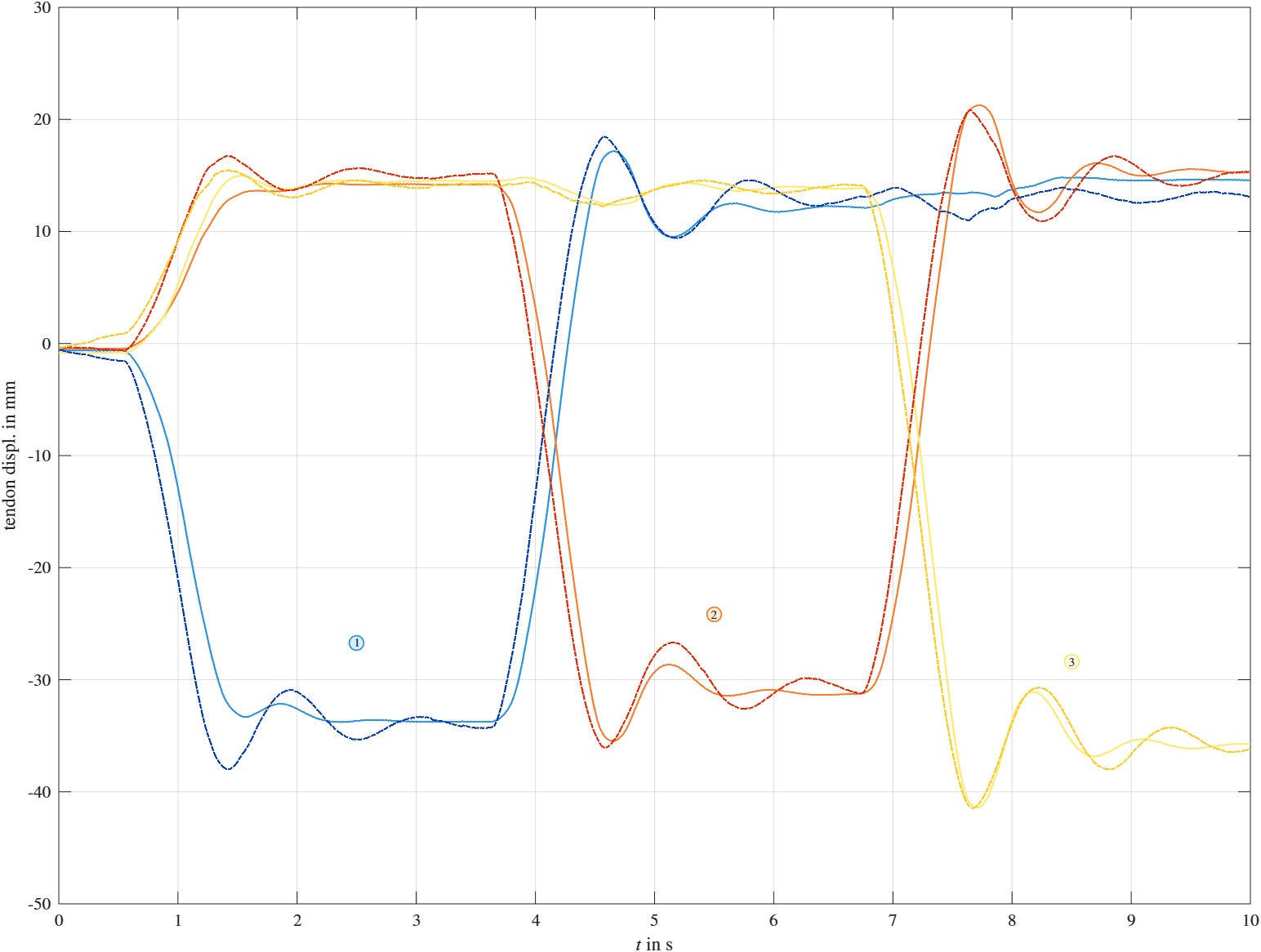}
      \vspace*{-0.5cm}
      \caption{Tendon displacements}
      \label{fig_val_model_dyn_L}
   \end{subfigure}
   \hfill
   \begin{subfigure}[t]{0.245\linewidth}
      \includegraphics[width=\linewidth]{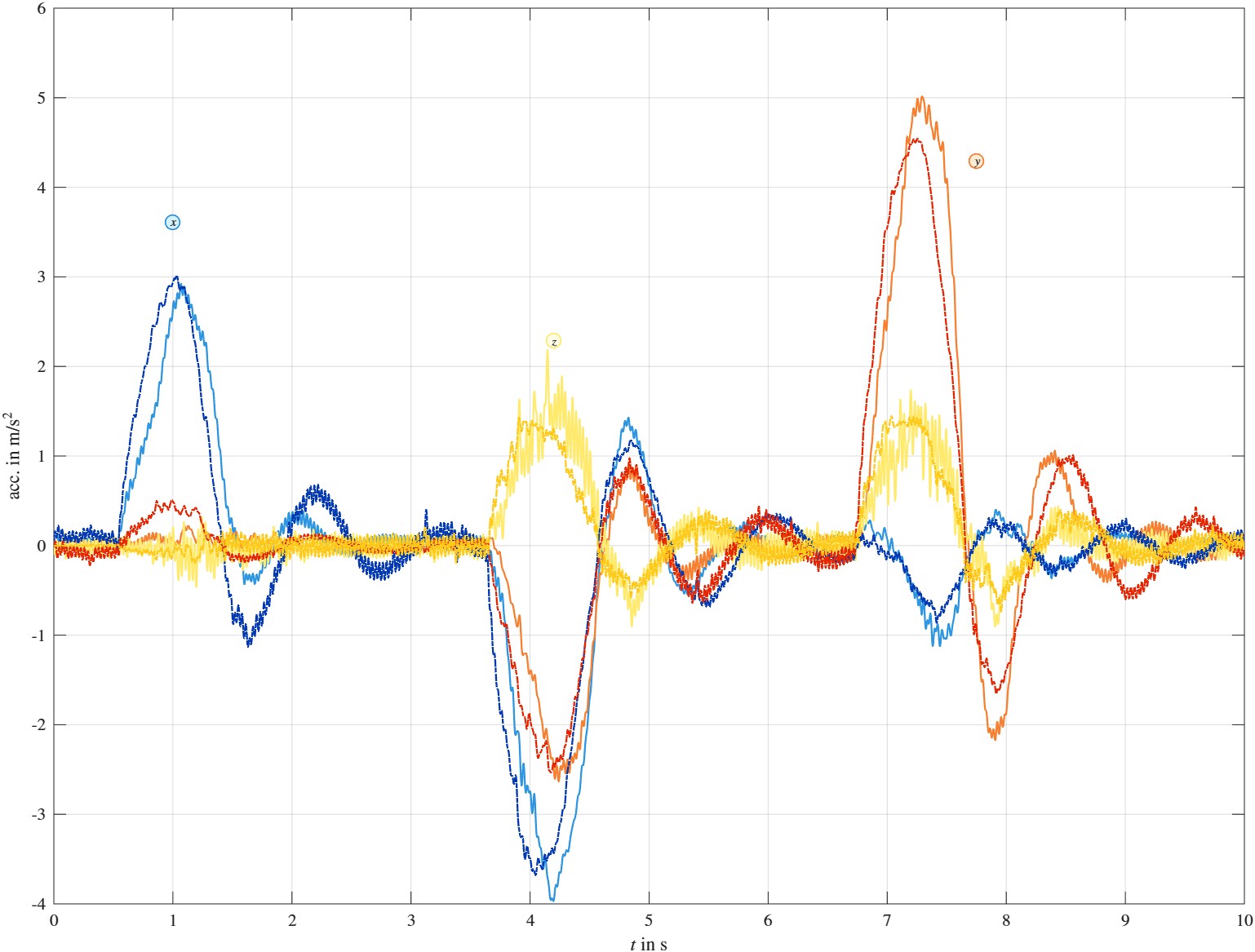}
      \vspace*{-0.5cm}
      \caption{Angular velocities of IMU 2}
      \label{fig_val_model_dyn_gyr}
   \end{subfigure}
   \hfill
   \begin{subfigure}[t]{0.245\linewidth}
      \includegraphics[width=\linewidth]{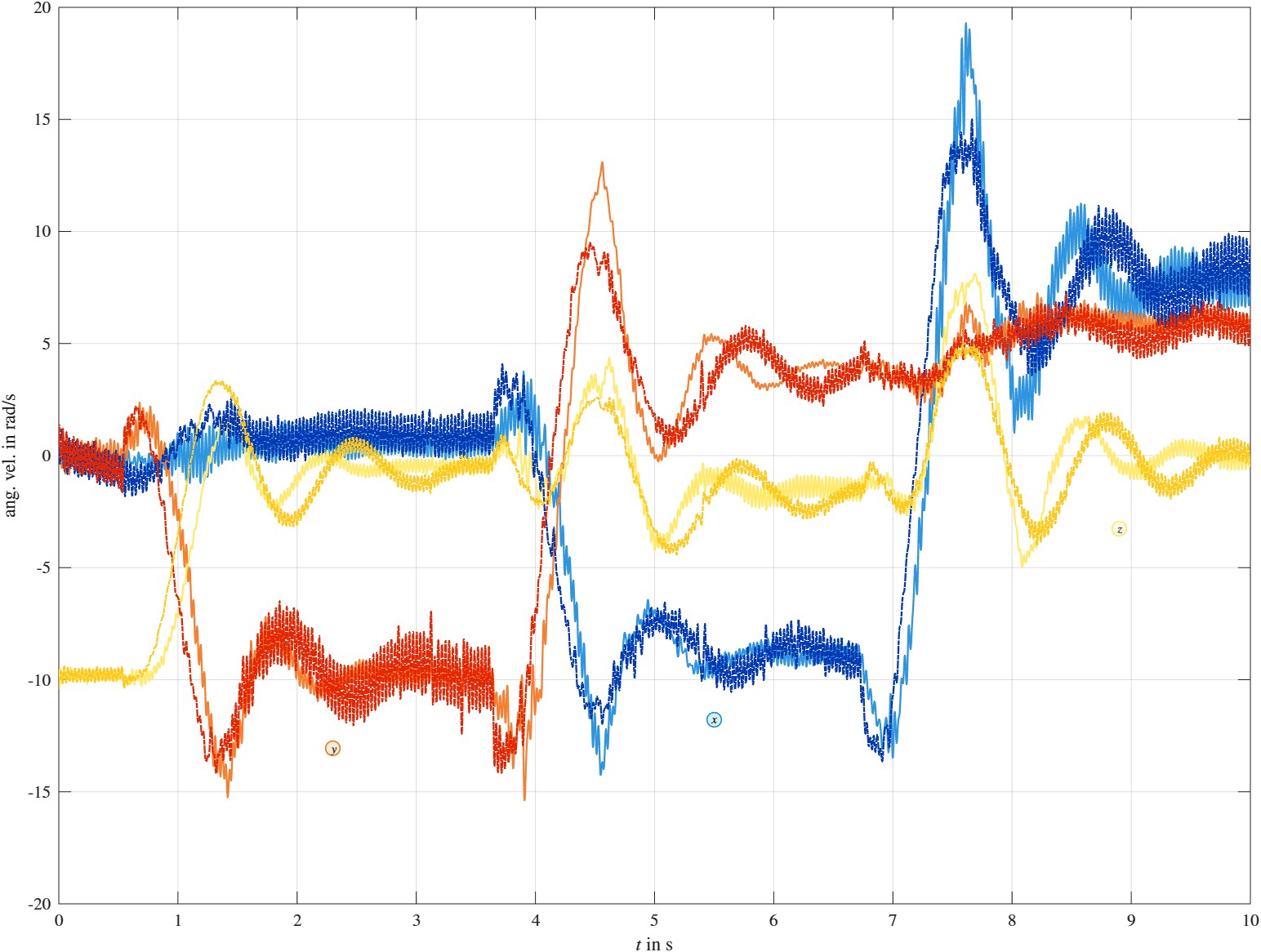}
      \vspace*{-0.5cm}
      \caption{Accelerations of IMU 2}
      \label{fig_val_model_dyn_acc}
   \end{subfigure}
   \caption{Results of the dynamic validation,
   where the system outputs during a dynamic motion are compared to simulated outputs.
   In (a), the bright colors denote the measured tensions (used as simulation inputs),
     the dark colors the prescribed reference tensions.
     In (b) to (d), the bright colors denote the experiment, the dark colors the simulation.}
   \label{fig_val_model_dyn}
\end{figure*}

For dynamic validation, we record the dynamic response of the experimental system to a dynamic input trajectory and compare the response to a simulation with the same inputs.
The commanded tendon tensions are shown in Fig.~\ref{fig_val_model_dyn} (a) together with the actually measured tensions; as the input for the simulation, the measured tensions are used.
Figs.~\ref{fig_val_model_dyn} (b) to (d) show the results of the dynamic motion for both the real system and the simulation.
The simulation captures the most important characteristics of the system response well, without any significant steady-state offsets in the tendon displacements and accelerations.
While only the results for the second IMU are shown, the data for the first IMU is very similar.
The \SI{10}{\second} dynamic simulation was done with a time step of $h = {2^{-9}}$~s $ \approx \SI{1.95e-03}{\second}$;
the computational time was \SI{0.38}{\second} (MATLAB R2025b with C\texttt{++} code generation).\footnote{For comparison, integrating the continuous-time equivalent of the presented model with standard, non-structure-preserving integrators for stiff systems takes much longer:
MATLAB \texttt{ode15s} takes \SI{30.07}{\second}, and \texttt{ode23t} \SI{27.18}{\second} (both with reduced tolerances \texttt{RelTol 1e-2} and \texttt{AbsTol 1e-3}),
indicating the excellent numerical efficiency of the proposed discrete model for time integration of the stiff TDCR.}

Finally, we visually compare the 3D backbone shape of a static simulation to the real robot; the visual matching is done by estimating the 3D pose of ArUco markers attached to the real robot.
Figs.~\ref{fig_val_model_static} (c) and (d) show the robot in two exemplary static configurations that correspond to constant tensions on two of the three straight tendons. The mismatch between backbone shape and simulation is only minimal and can be prospectively further optimized by adjusting the model parameters.

\subsection{Observer Validation}
\begin{figure}
   \centering
  \begin{subfigure}[t]{0.49\linewidth}
       \includegraphics[width=\linewidth]{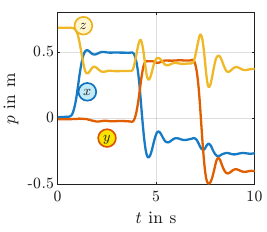}
       \vspace*{-0.4cm}
       {\small\caption{Without disturbance}}
       \label{fig:observed_tip_position_no_disturbance}
   \end{subfigure}
   \hfill
   \begin{subfigure}[t]{0.49\linewidth}
      \includegraphics[width=\linewidth]{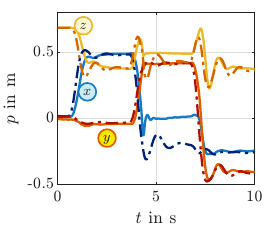}
      \vspace*{-0.4cm}
      \caption{With disturbance}
      \label{fig:observed_tip_position}
   \end{subfigure}
      \caption{Observed tip position with reference inputs.}
   \label{fig:observed_tip}
\end{figure}

\begin{figure}
   \centering
   \begin{subfigure}[t]{0.49\linewidth}
      \includegraphics[width=\linewidth]{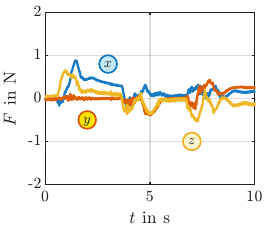}
      \vspace*{-0.4cm}
      \caption{Unconstrained environment}
      \label{fig:observed_disturbance_unconstrained}
   \end{subfigure}
   \hfill
   \begin{subfigure}[t]{0.49\linewidth}
      \includegraphics[width=\linewidth]{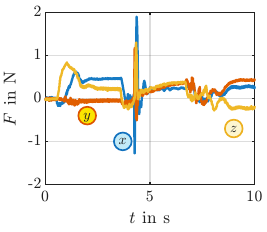}
      \vspace*{-0.4cm}
      \caption{Constrained environment}
      \label{fig:observed_disturbance_constrained}
   \end{subfigure}
   \caption{Observed disturbances with reference inputs.}
   \label{fig:observed_disturbance}
\end{figure}

We leverage the previously validated model to implement the combined state and disturbance observer with $n=8$ segments. The observer was tuned empirically.
For this configuration, we were able to achieve real-time performance for a sampling time of $h =  \SI{5}{\milli \second}$.
The reference input trajectories shown in Fig.~\ref{fig_val_model_dyn} (a) are used. Fig.~\ref{fig:observed_tip} (a) shows the resulting observed tip position without any environmental constraints. The observed disturbances are shown in Fig.~\ref{fig:observed_disturbance} (a).
As can be seen, the disturbances are bounded and remain below $ \SI{1}{\newton}$ and can be reasonably explained by model uncertainties such as friction or tendon elongation.

The effectiveness of the external disturbance term is demonstrated in Fig.~\ref{fig:observed_tip} (b), which shows the tip position under an imposed constraint. Between
$t\approx \SI{4}{\second}$ and $t\approx \SI{7}{\second}$ the manipulator is physically prevented from moving in the negative $x$-direction. The darker curve represents the observer output when only the state is estimated. In contrast, the extended-state observer accurately captures the imposed constraint, confirming its capability to identify and track external disturbances. The effect of the resulting force is further directly visible in Fig.~\ref{fig:observed_disturbance} (b).


\section{Conclusion}

We proposed a fully discrete, numerically efficient approach for modeling and estimation of continuum robots.
Using the same structure-preserving discretization framework, we extended the symplectic model of \cite{HK24} with tendon actuation for arbitrary tendon paths.
The model is then used as the basis for disturbance estimation, where we proposed an EKF with an extended state formulation that is able to track the manipulator's state in constrained and unconstrained environments, while providing an estimate of the disturbances acting on the manipulator.
Both the model and the observer were validated experimentally,
highlighting their accuracy and applicability to TDCR systems.

In future work, we plan to identify the model parameters directly from experimental data, conduct a more detailed validation of the model and the observer that also includes quantitative assessments,
and finally use both for more complex model-based control tasks.




\section*{DECLARATION OF GENERATIVE AI AND AI-ASSISTED TECHNOLOGIES IN THE WRITING PROCESS}
During the preparation of this work, the authors used ChatGPT for minor editing support. After using this tool, the authors reviewed and edited the content as needed and take full responsibility for the content of the publication.

\bibliography{library_ifac,literatur_forschung}             

\appendix
\end{document}